\documentclass[12pt]{article}
\usepackage{cite}
\usepackage{graphicx}
\usepackage{amssymb}
\def\qs{\langle \bar s s \rangle }
\def\qd{\langle \bar q q \rangle }

\begin{document}
\title{Mixing Angle of Hadrons in QCD: A New View}
\author{T. M. Aliev$^a$ \thanks{Permanent institute: Institute of Physics, Baku, Azerbaijan}
\thanks{taliev@metu.edu.tr},
A. Ozpineci$^a$ \thanks{ozpineci@metu.edu.tr},
~V. Zamiralov$^b$ \thanks{zamir@depni.sinp.msu.ru}
\\
{\small
$^a$Middle East Technical University, Ankara, Turkey} \\
{\small
$^b$Institute of Nuclear Physics, M. V. Lomonosov MSU, Moscow, Russia
}}
\begin{titlepage}
\maketitle
\thispagestyle{empty}
\begin{abstract}
A new method for calculation of the mixing angle between the hadrons within QCD sum rules is proposed.
In this method, the mixing is expressed in terms of quark and gluon degrees of freedom.
As an application, 
the detailed calculation of the mixing angle between heavy cascade
baryons $\Xi_Q$ and $\Xi_Q'$, $Q=c,~b$ is presented and it is found that the mixing angle between $\Xi_b$ ($\Xi_c$) and
$\Xi_b'$ ($\Xi_c'$) is given by $\theta_b = 6.4^\circ \pm 1.8 ^\circ$
($\theta_c = 5.5^\circ \pm 1.8^\circ$).
\vspace*{1cm}
\end{abstract}
\end{titlepage}

\section{Introduction}
During the last years, many new and unexpected experimental data 
appeared on heavy hadron spectroscopy \cite{R1}. The study of spectroscopy and decays of heavy hadrons can give essential 
information on the quark structure of these hadrons, in particular of the properties of the cascade baryons \cite{R2}.

The impressive progress on the experimental physics stimulated 
comprehensive theoretical studies for understanding dynamics of heavy flavored hadrons at the hadronic scale. This scale belong to 
the non-perturbative sector of QCD and therefore for the 
calculations of different characteristics of the baryons, a non-perturbative approach is needed. Among existing non-perturbative
methods, the QCD sum rules method \cite{R3} is one of the most predictive in 
studying properties of hadrons. The main ingredient of the QCD sum rules method is the choice of the interpolating
currents which is directly related to the quark content of the corresponding baryons.

Approximate flavor symmetries are useful tools in classifying hadrons. Breaking of these symmetries might lead to mixing of hadrons which differ
only in the flavor quantum numbers. For example, two baryons that mix due to flavor symmetry violation are $\Xi_Q$ and $\Xi'_Q$ baryons, both of which
are made up of $q$ ($q=u,~d$), $s$ and $Q$ ($Q=b,~c$) quarks. This mixing
of charmed $\Xi_c$ baryons has been calculated using quark model 
in \cite{franklin1,franklin2} and using heavy quark effective theory with $1/m_Q$ 
corrections, in \cite{ito}. In \cite{R2}, it was shown that this 
mixing can be important in determining
properties of these baryons and a framework has been proposed to calculate this mixing within the QCD sum rules method.
In the present work, we demonstrate the new approach for the calculation of the mixing angle between hadrons in the QCD sum rules framework and an application of the proposed method to study the 
mixing angle between the $\Xi_Q$ and $\Xi'_Q$ baryons carried out.

Approximate $SU(3)_f$ flavor symmetry of QCD allows us to identify the observed hadrons with multiplet of the $SU(3)_f$ group. If this symmetry were exact, the observed hadrons all would have definite flavor quantum
numbers under this flavor symmetry. The mass differences between the light $u$, $d$ and $s$ quarks, breaks this symmetry explicitly and hence the mass eigenstates need not be definite eigenstates of the $SU(3)_f$
symmetry, i.e. definite flavor eigenstates can mix to form the physically observed particles (unless there is another symmetry which prevents this mixing).

In QCD sum rules, the main object of interest is a correlation function of the form
\begin{eqnarray}
\Pi = i \int d^4x e^{i p x} \langle 0 \vert  {\cal T} \eta_H(x) \bar \eta_H \vert 0 \rangle
\label{corr}
\end{eqnarray}
where $\cal T$ is the time ordering operator and $\eta_H$ is an operator that can create the hadron $H$ from the vacuum.
If the pure  $H_1^0$ and $H_2^0$ states mix, 
then the physical states, i.e. states that have  definite mass, should be represented as a linear combination of these states. In such a case,
currents corresponding to  physical states should also be written as a superposition of the pure operators:
\begin{eqnarray}
\eta_{H_1} = \cos \theta \eta_{H_1^0} + \sin \theta \eta_{H_2^0}
\nonumber \\
\eta_{H_2} = -\sin \theta \eta_{H_1^0} + \cos \theta \eta_{H_2^0}
\end{eqnarray}
where $\theta$ is the mixing angle between $H_1$ and $H_2$.

Consider the following two-point correlation function:
\begin{eqnarray}
\Pi_{H_1 H_2}(p) = i \int d^4 x e^{i px} \langle 0 \vert {\cal T} \{ \eta_{H_1}(x) \eta_{H_2}(0) \} \vert 0 \rangle
\end{eqnarray}
If $\eta_{H_{1(2)}}$ creates only $H_{1(2)}$ and not the other one, this correlation function should be zero. Hence, the angle $\theta$ should be chosen in such a way as to give  $\Pi_{H_1H_2} = 0$. 

The general form of the correlation function can be written as
\begin{eqnarray}
\Pi_{H_1H_2}(p) = \not\!p \Pi_1(p^2) + \Pi_2(p^2)
\end{eqnarray}
Generally speaking, any of the functions $\Pi_i$ ($i=1,~2$) can be used for the sum rules analysis. In this work, the coefficient of the $\not\!p$ structure has been chosen to obtain a prediction for
the mixing angle.

Using the notation
\begin{eqnarray}
\Pi^0_{H_i H_j} = i \int d^4 x e^{i px} \langle 0 \vert {\cal T} \{ \eta_{H_1}(x) \eta_{H_2}(0) \} \vert 0 \rangle
\end{eqnarray} 
it is straightforward to show that the angle $\theta$ that makes $\Pi_{H_1 H_2} = 0$ can be expressed as:
\begin{eqnarray}
\tan 2 \theta = \frac{- a c \pm b \sqrt{b^2 + a^2 - c^2}}{- b c \mp a \sqrt{b^2 + a^2 - c^2}}
\end{eqnarray}
where $a = \frac12 ( \Pi^0_{H_2H_2} - \Pi^0_{H_1H_2} )$, $b= \frac12 (\Pi^0_{H_1 H_2}+\Pi^0_{H_2 H_1})$
and $c = \frac12 (\Pi^0_{H_1 H_2}- \Pi^0_{H_2 H_1})$. This expression can be simplified further by noting that, as explicit calculation have shown, 
$\Pi^0_{H_2 H_1} = \Pi^0_{H_1 H_2}$, i.e. $c=0$, which yields
\begin{eqnarray}
\tan 2 \theta = - \frac{b}{a} = \frac{2 \Pi^0_{H_2 H_1}}{\Pi^0_{H_1 H_1} - \Pi^0_{H_2 H_2}}
\label{thetaeq}
\label{tangenttheta}
\end{eqnarray}
Note that, Eq. \ref{thetaeq} has two solutions for 
$0^\circ < \theta < 180^\circ$ that differ by $90^\circ$.
In this work, we present the solution that is close to $0^\circ$. The other
solution corresponds to exchanging the identification of the dominant part
of $H_{1(2)}$ with $H_{2(1)}^0$. Further studies of other properties of $H_{1(2)}$ baryons is necessary
to remove this ambiguity.

After setting up the framework for the study of the mixing angle between the hadrons, let us concentrate on this mixing angle of the heavy cascade baryons.
In $SU(3)_f$ classification, baryons containing two light and one heavy quarks can be grouped into an anti-triplet and a sextet representation.
It is well known that the dominant components of $\Xi_Q$ and $\Xi'_Q$ belong to 
the anti-triplet and sextet representations of $SU(3)_f$ respectively, i.e. 
 $\Xi_Q(\Xi'_Q)$ is approximately antisymmetric(symmetric) under the exchange of light quarks. The reason of  why they are not exactly (anti)symmetric
 is the mixing between the $SU(3)_f$ representations.
 Note that, in the infinite heavy quark mass limit, there is an additional conserved quantity: the total angular momentum of the light degrees of freedom, $s_l$. 
 Since $\Xi^0_Q$ and ${\Xi'}^0_Q$ correspond to different values of $s_l$, this additional symmetry prevents the mixing of these two states.

Using the $SU(3)_f$ classification, the interpolating currents for the unmixed states can be chosen as:

\begin{eqnarray}
	\eta_{\Xi_Q}^0 = \frac{1}{\sqrt6} \epsilon^{abc} && \left[ 
		 2 (s_a^T C q_b) \gamma_5 Q_c + 2 t(s_a^T C \gamma_5 q_b) Q_c 
	\right. \nonumber \\
		&& + (s_a^T C Q_b) \gamma_5 q_c +  t(s_a^T C \gamma_5 Q_b) q_c
	\nonumber \\ 
		&& - \left. (q_a^T C Q_b) \gamma_5 s_c -  t(q_a^T C \gamma_5 Q_b) s_c \right]
\nonumber \\
	\eta_{\Xi'_Q}^0 = \frac{1}{\sqrt2} \epsilon^{abc} && \left[
	(s_a^T C Q_b) \gamma_5 q_c +  t(s_a^T C \gamma_5 Q_b) q_c +
	\right. \nonumber \\
	&& \left. (q_a^T C Q_b) \gamma_5 s_c  + t(q_a^T C \gamma_5 Q_b) s_c \right]
\end{eqnarray}
Here $a,~b$ and $c$ are the color indices, $q=u$ or $d$ and $t$ is an arbitrary auxiliary parameter. The $t=-1$ case, corresponds to the Ioffe current.
The notation $\eta^0$ is used to denote that these operators are pure $\bar 3$ or $6$ operators.

The correlation function, Eq. \ref{corr}, can also be calculated in terms of quark and gluon degrees of freedom in 
the deep Euclidean region, $p^2 << 0$, using operator product expansion.
The analytical results for the corresponding correlators are presented in the appendix.

\section{Numerical Analysis}
In this section, we present the numerical analysis of our results. The
numerical values for the input parameters are $\qd = (-0.243~GeV)^3$,
$\frac{\qs}{\qd} = 0.8$, $\langle g_s^2 G^2 \rangle = 0.47~GeV^4$,
$m_0^2=0.8~GeV^2$, $m_b = 4.8~GeV$, $m_c = 1.4~GeV$ and $m_s=0.14~GeV$.

The sum rules results depend on three auxiliary parameters: the
continuum threshold $s_0$, the Borel parameter $M^2$ and the arbitrary
parameter $t$ in the interpolating current. The continuum threshold 
should be close to the first excited state that can couple to the current.
In our analysis, it is chosen to be near $s_0 = (m_B + 0.5)^2$.
The lower limit of the working region of the Borel parameter can be found by requiring that the contribution of the highest dimensional
operator is less than $20\%$ of the perturbative term.
Its upper limit is determined in such a way that the contribution of the
higher states and the continuum is less then the contribution of the first
pole.

In Figs. 1(2), the dependence of $\theta_b$($\theta_c$) 
on the Borel parameter
is plotted, at $t=3$. For $\Xi_b$, the continuum threshold is chosen to
be $s_0=40~GeV^2$, and $42~GeV^2$, and for $\Xi_c$, it is chosen to be
$s_0=9~GeV^2$ and $10~GeV^2$. It is seen that although the
Borel parameter is allowed to change in a wide range of values, there is
practically no dependence of the predictions of the sum rules 
on the value of the
Borel parameter.

In Figs. 3(4), the mixing angle $\theta_b$($\theta_c$) is plotted
as function of $\cos \alpha$, where $\alpha$ is defined through
$t=tan\alpha$ at $M^2=10~GeV^2$ and $15~GeV^2$($M^2=5$ and $8~GeV^2$). It is seen that in both graphs, the mixing angle take very
large values at certain values of
$\cos \alpha$. This enhancement can be understood by noting that the
tangent of the mixing angle, Eq. \ref{tangenttheta}, is a ratio of two sum
rules. In principal, both the numerator and the denominator should become
zero at the same value of $t$. But due to approximations in the sum rules,
their zeros are shifted. Hence, when the denominator becomes zero, and the
numerator is non-zero (although small), $\tan \theta$ diverges. Near these
points the sum rules is not reliable, hence in obtaining a prediction for
the mixing angle, one should keep away from these points. From the figures,
it is seen that, sum rule predictions on the mixing angle is almost
independent of the value of $t$ chosen.

Finally, the predictions on the mixing angle between $\Xi_Q$ and
$\Xi_Q'$ baryons of QCD sum rules are: 
\begin{eqnarray}
\theta_b = 6.4^\circ \pm 1.8^\circ
\end{eqnarray}
for the $\Xi_b-\Xi_b'$ mixing and
\begin{eqnarray}
\theta_c = 5.5^\circ \pm 1.8^\circ
\end{eqnarray}
for the $\Xi_c-\Xi_c'$ mixing. In these predictions, the uncertainties are
due to the neglected higher dimensional operators and the uncertainties in
the auxiliary parameters of the sum rules. The largest source of uncertainty
is the variations in the continuum threshold $s_0$. In Table \ref{table},
sum rules predictions and the predictions of \cite{franklin1,franklin2,ito} are
presented. It is seen that, within errors, the predictions for $\theta_c$
 are in agreement, whereas, our prediction on $\theta_b$ is in 
 disagreement with the prediction of \cite{franklin2}.

 As also mentioned previously, in the heavy quark limit, the mixing angle should be zero.
 The heavy quark limit of the sum rules can be obtained by taking the $m_Q \rightarrow \infty$ limit after setting 
$M^2 \rightarrow m_Q^2 + 2 m_Q T$, $s \rightarrow m_Q^2 + 2 m_Q \nu$ and $s_0 \rightarrow m_Q^2 + 2 m_Q \nu_0$, where $T$ is the new Borel parameter, $\nu$ is the four velocity of the heavy baryon, and $\nu_0$ is the threshold in the heavy quark limit.
Using the expressions in the
appendix, it can be shown that, in the heavy quark limit $\tan \theta_Q \propto \frac{1}{m_Q^2}$.
But the numeric results for the $b$ and $c$ quarks show 
that the suppression of the mixing angle in the heavy quark limit is not realized at the physical $b$ quark mass.
The suppression starts at much larger values of the heavy quark mass.

\begin{table}[h!]
\begin{center}
\begin{tabular}{|c|c|c|c|c|}
\hline
$\theta_Q$ & This Work & \cite{franklin1}  &\cite{franklin2} & \cite{ito} \\
\hline
$\theta_c$ & $5.5^\circ\pm1.8^\circ$ & $3.8^\circ$ & $3.8^\circ$& $14^\circ\pm14^\circ$\\
\hline
$\theta_b$ & $6.4^\circ\pm1.8^\circ$ & -- & $1.0^\circ$ & -- \\
\hline
\end{tabular}
\caption{The predictions of QCD sum rules on the mixing angle along with
the prediction of quark model \cite{franklin1,franklin2} and heavy quark effective
theory \cite{ito}}
\end{center}
\label{table}
\end{table}

In conclusion, a new method is presented for the determination of the mixing angle between the hadrons in QCD sum rules. This method is applied to the calculation of the mixing angle
between the heavy cascade hyperons $\Xi_Q$ and $\Xi'_Q$.
\section*{Acknowledgment}
One of the authors (V.Z.) is grateful to B. L. Ioffe for useful discussion at the early stages of the work.

\appendix

\section{Analytical Results}
In this appendix, we present the explicit expression for the coefficient of the $\not\!p$ structure in the correlation function

\begin{eqnarray}
\Pi^0_{\Xi \Xi}  &=& 
- \frac{1}{288 M^4} e^{-\frac{m_Q^2}{M^2}} m_0^2 \qd \qs m_Q \left(m_s (-1+t)^2 + 2 m_Q (-13 + 2 t + 11 t) \right) 
\nonumber \\
&&  + \frac{1}{144 M^2} e^{- \frac{m_Q^2}{M^2}} m_s m_Q \qd \qs (1 + 4 t - 5 t^2)
\left( 1 - \frac{5}{12} \frac{m_Q^2 m_0^2}{M^4} \right)
\nonumber \\
&& + \frac{1}{288 M^2} e^{- \frac{m_Q^2}{M^2}} m_0^2 \qd \qs (25 - 2 t - 23 t^2)
\nonumber \\
&& - \frac{1}{72} e^{- \frac{m_Q^2}{M^2}} \qd \qs ( 13 - 2 t - 11 t^2)
\nonumber \\
&& + \frac{1}{768 \pi^2} e^{- \frac{m_Q^2}{M^2}} m_s m_0^2 \left( 3 \qs (1+t)^2 + 2 \qd (-13 + 2 t + 11 t^2) \right)
\nonumber \\
&& + \int_{m_Q^2}^\infty ds e^{- \frac{s}{M^2}} \rho_{11}(s)
\end{eqnarray}
where
\begin{eqnarray}
\rho_{11}(s) &=& 
\frac{3 m_Q^4}{512 \pi^4}  (5 + 2 t + 5 t^2) \ln (1 + \hat s) 
\nonumber \\
&& + \frac{3 m_Q^3}{384\pi^2}  \hat s \frac{(2 + \hat s)}{(1+ \hat s)^2} 
\left[ \frac{m_Q}{16\pi^2}(-6 -6 \hat s + \hat s^2) + \frac{m_s}{m_Q^3} \qs \right](5 + 2 t + 5 t^2)
\nonumber \\
&& + \frac{1}{1536 \pi^4} m_Q^3 m_s (-1 - 4 t + 5 t^2) \left[ 6 \ln (1 + \hat s) - \hat s \frac{6 + 9 \hat s + 2 \hat s^2}{(1+\hat s)^2} \right]
\nonumber \\
&& - \frac{1}{192 \pi^2} m_Q (\qd + \qs) \frac{\hat s^2}{(1 + \hat s)^2} (-1 -4 t + 5 t^2)
\nonumber \\
&&  - \frac{1}{192 \pi^2} m_s \hat s \frac{2 + \hat s}{(1 + \hat s)^2} \qd (-13 + 2 t + 11 t^2) 
\nonumber \\
&& -  \frac{1}{768 m_Q \pi^2} m_0^2 (\qd + \qs) \frac{1}{(1+\hat s)^2} (-1+t) (-1 - 5 t + 6 \hat s(1 + t))
\nonumber \\
&& +  \frac{1}{768 m_Q^2 \pi^2} m_s m_0^2 (\qd + \qs) \frac{1}{(1 + \hat s)^2}(-1+t)^2 
\nonumber \\
&& + \frac{1}{16 m_Q^2 \pi^2} m_s m_0^2 \qd \frac{1}{(1 + \hat s)^3} (1-t^2) \ln \hat s 
\nonumber \\
&& -  \frac{1}{32 \pi^2 M^2} m_s m_0^2 \qd (-1+t^2) \left[ \ln \frac{m_Q^2}{\Lambda^2} + \left(1 + \frac{1}{(1+\hat s)^2}\right)\ln \hat s \right]
\end{eqnarray}

\begin{eqnarray}
\Pi^0_{\Xi' \Xi'} &=& 
 - \frac{1}{48 M^4} e^{-\frac{m_Q^2}{M^2}} m_0^2 \qd \qs m_Q (-1+t) \left(m_s (1+t) +  m_Q (-1 +  t) \right) 
\nonumber \\
&&  + \frac{1}{16 M^2} e^{- \frac{m_Q^2}{M^2}} m_s m_Q \qd \qs (1  -  t^2)
\left( 1 - \frac{5}{12} \frac{m_Q^2 m_0^2}{M^4} \right)
\nonumber \\
&& - \frac{1}{96 M^2} e^{- \frac{m_Q^2}{M^2}} m_0^2 \qd \qs (-1 + t)^2
\nonumber \\
&& + \frac{1}{24} e^{- \frac{m_Q^2}{M^2}} \qd \qs ( -1 +t)^2
\nonumber \\
&& + \frac{1}{768 \pi^2} e^{- \frac{m_Q^2}{M^2}} m_s m_0^2 \left( - \qs (13+10 t+13t^2) + 6 \qd (-1+t)^2) \right)
\nonumber \\
&& + \int_{m_Q^2}^\infty ds e^{- \frac{s}{M^2}} \rho_{22}(s)
\end{eqnarray}
where
\begin{eqnarray}
\rho_{22}(s) &=& 
\frac{3 m_Q^4}{512 \pi^4}  (5 + 2 t + 5 t^2) \ln (1 + \hat s) 
\nonumber \\
&& + \frac{m_Q^3}{128\pi^2}  \hat s \frac{(2 + \hat s)}{(1+ \hat s)^2} 
\left[\frac{m_Q}{16\pi^2} (-6 -6 \hat s + \hat s^2) + \frac{m_s}{m_Q^3} \qs \right] (5 + 2 t + 5 t^2)
\nonumber \\
&& + \frac{3}{512 \pi^4} m_Q^3 m_s (-1  +  t^2) \left[ 6 \ln (1 + \hat s) - \hat s \frac{6 + 9 \hat s + 2 \hat s^2}{(1+\hat s)^2} \right]
\nonumber \\
&& - \frac{3}{64 \pi^2} m_Q (\qd + \qs) \frac{\hat s^2}{(1 + \hat s)^2} (-1  +  t^2)
\nonumber \\
&& -   \frac{1}{64 \pi^2} m_s \hat s \frac{2 +  \hat s}{(1 + \hat s)^2} \qd (-1+t)^2 
\nonumber \\
&& -  \frac{1}{256 m_Q \pi^2} m_0^2 (\qd + \qs) \frac{6 \hat s-7}{(1+\hat s)^2} (-1+t^2) 
\nonumber \\
&& + \frac{1}{256 m_Q^2 \pi^2} m_s m_0^2 \frac{1}{(1 + \hat s)^2}\left[ - \qd (-1+t)^2 + \qs  (3 + 2 t + 3 t^2) \right]
\nonumber \\
\end{eqnarray}

\begin{eqnarray}
\sqrt3 \Pi^0_{\Xi \Xi'} = \sqrt3 \Pi^0_{\Xi' \Xi} &=&  
- \frac{1}{192 M^4} e^{-\frac{m_Q^2}{M^2}} m_s m_Q m_0^2 \qd \qs (-3 + 2 t +t ^2) 
\nonumber \\
&&  - \frac{1}{24 M^2} e^{- \frac{m_Q^2}{M^2}} m_s m_Q \qd \qs (-2  +  t + t^2)
\left( 1 - \frac{5}{12} \frac{m_Q^2 m_0^2}{M^4} \right)
\nonumber \\
&& + \int_{m_Q^2}^\infty ds e^{- \frac{s}{M^2}} \rho_{12}(s)
\end{eqnarray}
where
\begin{eqnarray}
\rho_{12}(s) &=& 
 - \frac{1}{256 \pi^4} m_Q^3 m_s (-2 + t +  t^2) \left[ 6 \ln (1 + \hat s) - \hat s \frac{6 + 9 \hat s + 2 \hat s^2}{(1+\hat s)^2} \right]
\nonumber \\
&& - \frac{1}{32 \pi^2} m_Q (\qd - \qs) \frac{\hat s^2}{(1 + \hat s)^2} (-2 + t +  t^2)
\nonumber \\
&& -  \frac{1}{128 m_Q \pi^2} m_0^2 (\qd - \qs) \frac{1}{(1+\hat s)^2} (-1+t) (-3 - 2 t + 3 \hat s(1 + t))
\nonumber \\
&& -  \frac{1}{128 m_Q^2 \pi^2} m_s m_0^2  \frac{1}{(1 + \hat s)^2}\left( \qd (-1+t^2) + 2 \qs (1+t+t^2) \right) 
\end{eqnarray}

where $\hat s = \frac{s}{m_Q^2} - 1$

The contribution of the higher states and the continuum is subtracted using quark hadron duality. It amounts to replacing the upper limit of integration in $s$ with $s_0$, i.e.
\begin{eqnarray}
\int_{m_Q^2}^\infty ds \rightarrow \int_{m_Q^2}^{s_0} ds
\end{eqnarray}

\newpage
\begin{figure}
\begin{center}
\includegraphics[width=9cm]{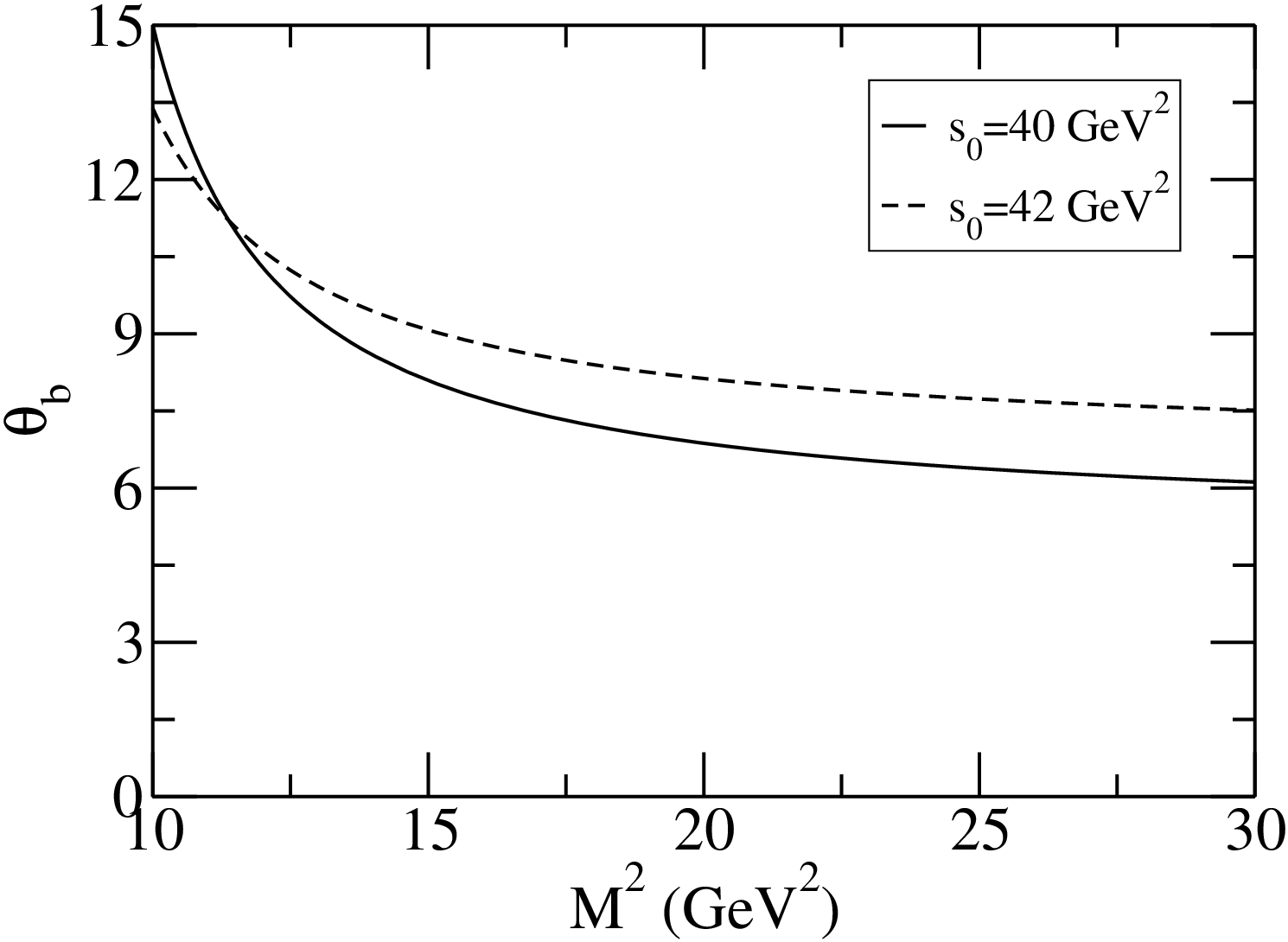}
\caption{The dependence of $\theta_b$ on the Borel parameter for $t=-3$ and
for two different values of the continuum threshold}
\end{center}
\end{figure}
\begin{figure}
\begin{center}
\includegraphics[width=9cm]{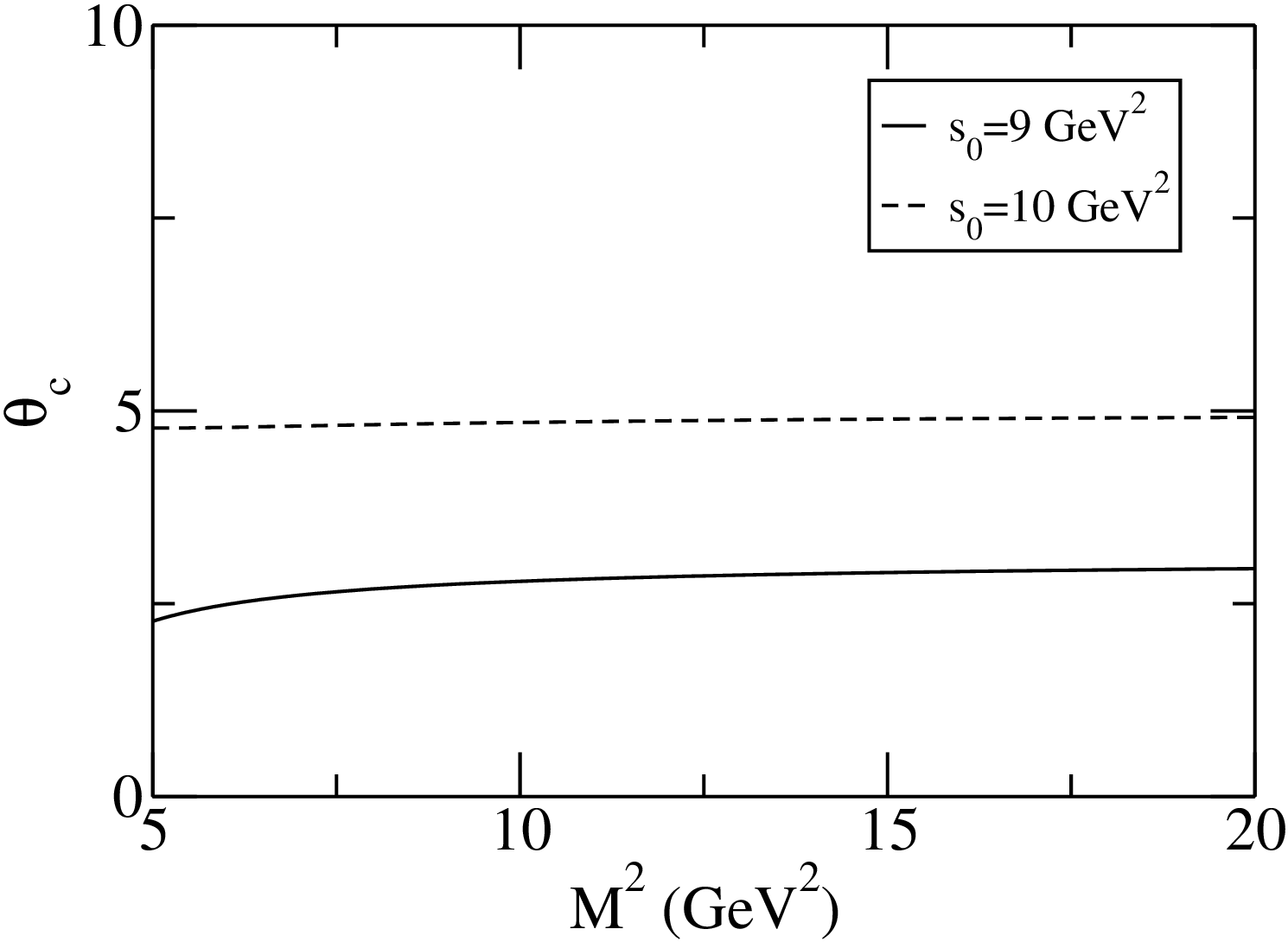}
\end{center}
\caption{
The same as Fig. 1, but for $\theta_c$}
\end{figure}
\begin{figure}
\begin{center}
\includegraphics[width=9cm]{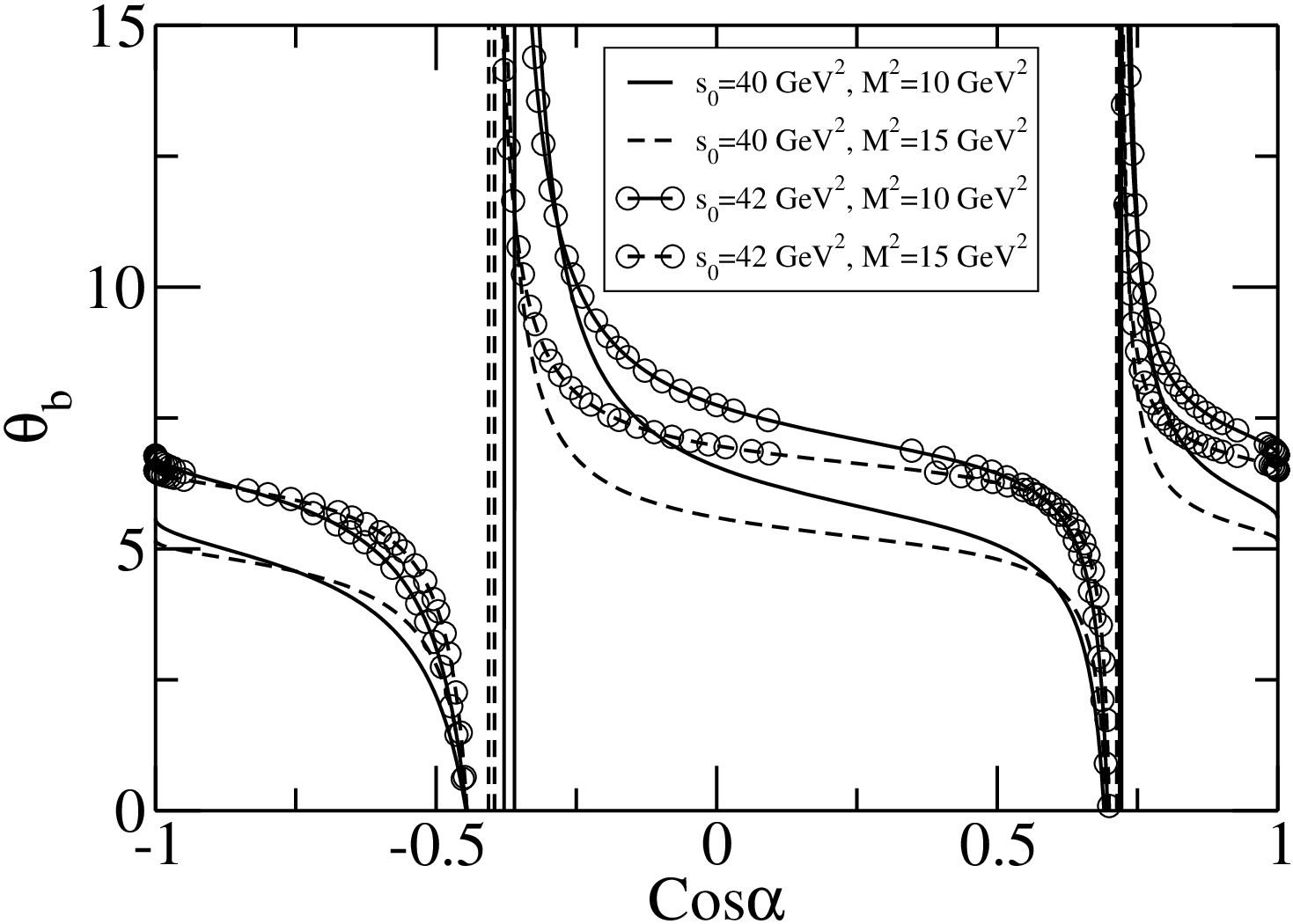}
\end{center}
\caption{The dependence of $\theta_b$ on $\cos \alpha$, where
$t=\tan\alpha$, for two different values of the continuum threshold, $s_0$
and the Borel parameter $M^2$}
\end{figure}

\begin{figure}
\begin{center}
\includegraphics[width=9cm]{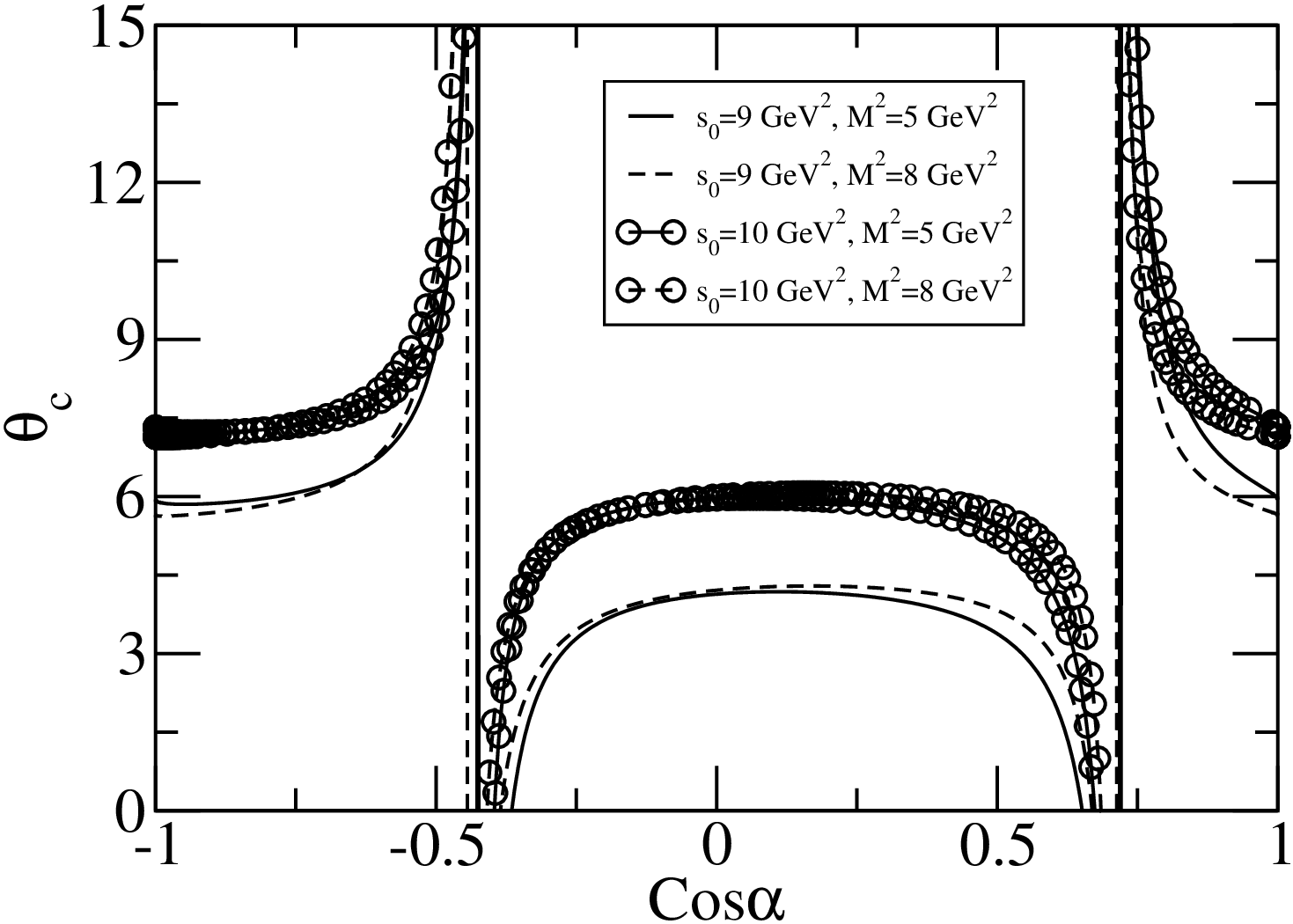}
\end{center}
\caption{
The same as Fig. 3, but for $\theta_c$}
\end{figure}
\end{document}